\documentclass[graybox]{svmult}

\usepackage{mathptmx}       
\usepackage{helvet}         
\usepackage{courier}        
\usepackage{graphicx}        
\usepackage{amsmath,amssymb,bm}
\usepackage{url}
\usepackage{isabelle,isabellesym}
\begin{document}


\title*{Formalising Mathematics In Simple Type Theory}
\author{Lawrence C. Paulson}
\institute{Computer Laboratory, University of Cambridge, England, \email{lp15@cam.ac.uk}}


\maketitle

\abstract{Despite the considerable interest in new dependent type theories, simple type theory \cite{church40} (which dates from 1940) is sufficient to formalise serious topics in mathematics. This point is seen by examining formal proofs of a theorem about stereographic projections. A formalisation using the HOL Light proof assistant is contrasted with one using Isabelle/HOL\@. Harrison's technique for formalising Euclidean spaces \cite{harrison-euclidean} is contrasted with an approach using Isabelle/HOL's axiomatic type classes~\cite{hoelzl-filters}. However, every formal system can be outgrown, and mathematics should be formalised with a view that it will eventually migrate to a new formalism.}

\section{Introduction}
\label{sec:intro}

Let's begin with Dana Scott:
\begin{quote}
No matter how much wishful thinking we do, the theory of types is here to stay. There is \textit{no other way} to make sense of the foundations of mathematics. Russell (with the help of Ramsey) had the right idea, and Curry and Quine are very lucky that their unmotivated formalistic systems are not inconsistent.\footnote{Italics in original}	\cite[p.\ts413]{scott93}
\end{quote}

The foundations of mathematics is commonly understood as referring to philosophical conceptions such as logicism (mathematics reduced to logic), formalism (mathematics as ``a combinatorial game played with the primitive symbols'')~\cite[p.\ts62]{neumann-foundations}, Platonism (``mathematics describes a non-sensual reality, which exists independently \dots{} of the human mind'') \cite[p.\ts323]{goedel-basic-foundations} and intuitionism (mathematics as ``a production of the human mind'') \cite[p.\ts52]{heyting-foundations}. Some of these conceptions, such as logicism and formalism, naturally lend themselves to the idea of doing mathematics in a formal deductive system. Whitehead and Russell's magnum opus, \textit{Principia Mathematica} \cite{principia}, is the quintessential example of this. Other conceptions are hostile to formalisation. However, a tremendous amount of mathematics has been formalised in recent years, and this work is largely indifferent to those philosophical debates.

This article is chiefly concerned with the great body of analysis and topology formalised by John Harrison, using higher-order logic as implemented in his HOL Light proof assistant~\cite{hol-light-tutorial}. The original motive for this work was to verify implementations of computer arithmetic, such as the calculation of the exponential function~\cite{harrison-exp}, prompted by the 1994 floating-point division bug that forced Intel to recall millions of Pentium chips at a cost of \$475 million~\cite{nicely-pentium-fdiv}. Another great body of mathematics was formalised by Georges Gonthier using Coq: the four colour theorem~\cite{gonthier-4ct}, and later, the odd order theorem~\cite{gonthier-oot}. Here the motive was to increase confidence in the proofs: the first four colour proof involved thousands of cases checked by a computer program, while the proof of the odd order theorem originally appeared as a 255-page journal article. Finally there was the Flyspeck project, to formalise Thomas Hales's proof of the Kepler conjecture, another gigantic case analysis; this formalisation task was carried out by many collaborators using HOL Light and Isabelle/HOL, so again higher-order logic.

Higher-order logic is based on the work of Church~\cite{church40}, which can be seen as a simplified version of the type theory of Whitehead and Russell. But while they were exponents of logicism, today's HOL Light and Isabelle/HOL users clearly aren't, or at least, keep their views secret.

Superficially, Coq users are indeed exponents of intuitionism: they regularly refer to constructive proofs and stress their rejection of the excluded middle. However, this sort of discussion is not always convincing. For example, the abstract announcing the Coq proof of the odd order theorem declares ``the formalized proof is constructive'' \cite[p.\ts163]{gonthier-oot}. This theorem states that every finite group of odd order is solvable, and therefore a constructive proof should provide, for a given group~$G$ of odd order, evidence that $G$ is solvable. However, the solvability of a finite group can be checked in finite time, so no evidence is required. So does the constructive nature of the proof embody anything significant? It turns out that some results in the theory of group modules could only be proved in double-negation form \cite[p.\ts174]{gonthier-oot}.

Analysis changes everything. Constructive analysis looks utterly different from classical analysis. As formulated by Bishop~\cite{bishop-bridges}, we may not assume that a real number~$x$ satisfies $x<0 \lor x=0 \lor x>0$ , and $x\not=0$ does not guarantee that $xy=1$ for some real~$y$. In their Coquelicot analysis library, Boldo et al.\ assume these classical principles, while resisting the temptation to embrace classical logic in full \cite[\S3.2]{boldo-coquelicot}.  

The sort of constructivism just described therefore seems to lack an overarching philosophical basis or justification. In contrast, Martin-L\"of's type theory was intended from the start to support Bishop-style constructive analysis \cite{martin-lof-itt-predicative}; this formal calculus directly embodies Heyting's intuitionistic interpretation of the logical constants~\cite{martin-lof-meanings}. It is implemented as the Agda \cite{bove-agda} programming language and proof assistant. 

It's worth remarking that the very idea of fixing a formalism as the \textit{foundation} of intuitionistic mathematics represents a sharp deviation from its original conception. As Heyting wrote,
\begin{quote}
	 The intuitionistic mathematician \dots{} uses language, both natural and formalised, only for communicating thoughts, i.e., to get others or himself to follow his own mathematical ideas. Such a linguistic accompaniment is not a representation of mathematics; still less is it mathematics itself.\cite[p.\ts52--3]{heyting-foundations}
\end{quote}

Constructive logic is well supported on the computer. However, the choice of proof assistant is frequently dictated by other considerations, including institutional expectations, the availability of local expertise and the need for specific libraries. The popularity of Coq in France is no reason to imagine that intuitionism is the dominant philosophy there.

Someone wishing to formalise mathematics today has three main options:
\begin{itemize}
  \item Higher-order logic (also known as simple type theory), where types are built inductively from certain base types, and variables have fixed types. Generalising this system through polymorphism adds considerable additional expressiveness.
  \item Dependent type theories, where types are parameterised by terms, embodying the propositions-as-types principle. This approach was first realised in NG de Bruijn's AUTOMATH~\cite{debruijn-survey}. Such systems are frequently but not necessarily constructive: AUTOMATH was mainly used to formalise classical mathematics.
  \item Set theories can be extremely expressive. The Mizar system has demonstrated that set theory can be a foundation for mathematics in practice as well as in theory~\cite{bancerek-lattices}.  Recent work by Zhan~\cite{zhan-fundamental} confirms this point independently, with a high degree of automation.
\end{itemize}
 All three options have merits. While this paper focuses on higher-order logic, I make no claim that this formalism is the best foundation for mathematics. It is certainly less expressive than the other two. And a mathematician can burst free of any formalism as quickly as you can say ``the category of all sets''. I would prefer to see a situation where formalised mathematics could be made portable: where proofs could be migrated from one formal system to another through a translation process that respects the structure of the proof.

\section{Higher-Order Logic on the Computer}
\label{sec:stt}

A succinct way to describe higher-order logic is as a predicate calculus with simple types, including functions and sets, the latter seen as truth-valued functions.

Logical types evolved rapidly during the 20th century. For Whitehead and Russell, types were a device to forestall the paradoxes, in particular by enforcing the distinction between sets and individuals. But they had no notation for types and never wrote them in formulas. They even proved (the modern equivalent of) $V\in V$, concealing the type symbols that prevent Russell's paradox here~\cite{feferman-typical-ambiguity}. Their omission of type symbols, which they termed \textit{typical ambiguity}, was a precursor to today's polymorphism. It seems that they preferred to keep types out of sight. 

Church \cite{church40} provided a type notation including a type~$\iota$ of individuals and a separate type $\omicron$ of truth values, with which one could express sets of individuals (having type $\omicron\iota$), sets of sets of individuals (type $\omicron(\omicron\iota)$) etc., analogously to the cumulative hierarchy of sets, but only to finite levels. Church assigned all individuals the same type.

Other people wanted to give types a much more prominent role. The mathematician NG de Bruijn devoted much of his later career, starting in the 1960s, to developing type theories for mathematics:
\begin{quote}
	I believe that thinking in terms of \textit{types} and \textit{typed sets} is much more natural than appealing to untyped set theory. \dots{}
	In our mathematical culture we have learned to keep things apart. If we have a rational number and a set of points in the Euclidean plane, we cannot even imagine what it means to form the intersection. The idea that both might have been coded in ZF with a coding so crazy that the intersection is \textit{not empty} seems to be ridiculous. If we think of a set of objects, we usually think of collecting things of a certain type, and set-theoretical operations are to be carried out inside that type. Some types might be considered as subtypes of some other types, but in other cases two different types have nothing to do with each other. That does not mean that their intersection is empty, but that it would be insane to even \textit{talk} about the intersection. \cite[p.\ts31]{debruijn-types}\footnote{Italics in original}
\end{quote}
De Bruijn also made the case for polymorphism:
\begin{quote}
Is there the drawback that working with typed sets is much less economic then with untyped ones? If things have been said for sets of apples, and if these same things hold, \textit{mutatis mutandis}, for sets of pears, does one have to repeat all what had been said before? No. One just takes a type variable, $\xi$ say, and expresses all those generalities for sets of things of type~$\xi$. Later one can apply all this by means of a single instantiation, replacing $\xi$ either by \textit{apple} or by \textit{pear}. \cite[p.\ts31]{debruijn-types}
\end{quote}
His work included the first computer implementations of dependent type theories. However, his view that apples and pears should have different types, using type variables to prevent duplication, is universally accepted even with simple type theory.

\subsection{Why Simple Type Theory?}

What is the point of choosing simple type theory when powerful dependent type theories exist? One reason is that so much can be done with so little. HOL Light ``sets a very exacting standard of correctness'' and ``compared with other HOL systems, \dots{} uses a much simpler logical core.''%
\footnote{\url{http://www.cl.cam.ac.uk/~jrh13/hol-light/}} 
Thanks to this simplicity, fully verified implementations now appear to be in reach \cite{kumar-self-formalisation}. Isabelle/HOL's logical core is larger, but nevertheless, concepts such as quotient constructions \cite{kaliszyk-quotients}, inductive and coinductive definitions \cite{blanchette-datatypes,paulson-coind}, recursion, pattern-matching and termination checking \cite{krauss-partial-recursive} are derived from Church's original HOL axioms; with dependent type theories, such features are generally provided by extending the calculus itself~\cite{gimenez-recursive}.

The other reason concerns automation. Derivations in formal calculi are extremely long. Whitehead and Russell needed hundreds of pages to prove 1+1=2 \cite[p.\ts360]{principia}.%
\footnote{In fact the relevant proposition,$*54\cdot 43$, is a statement about sets.  Many of the propositions laboriously worked out here are elementary identities that are trivial to prove with modern automation.}
Proof assistants must be capable of performing lengthy deductions automatically. But more expressive formalisms are more difficult to automate.  
Even at the most basic level, technical features of constructive type theories interfere with automation. \textit{Term rewriting} refers to the use of a set of identities to perform algebraic simplification. It has been a staple of automated theorem proving since the 1970s~\cite{bm79}. Isabelle/HOL has over 2800 rewrite rules pre-installed, and the full battery can be applied with the single word \textbf{auto}. The rewriting tactics of Coq \cite[\S8.6]{coq-refman} --- the most advanced implementation of dependent types --- apply a single, explicitly named, rewrite rule.  Recent versions of the Lean proof assistant (which implements the same calculus as Coq) finally provide strong simplification~\cite{avigad-lean}. 

It is also striking to consider the extent to which the Ssreflect proof language and library has superseded the standard Coq libraries. Gonthier and Mahboubi write
\begin{quote}
	Small-scale reflection is a formal proof methodology based on the pervasive use of computation with symbolic representations.  \dots{} The statements of many top-level lemmas,  and of most proof subgoals, explicitly contain symbolic representations; translation  between  logical  and  symbolic  representations  is  performed  under  the explicit, fine-grained control of the proof script. The efficiency of small-scale reflection hinges on the fact that fixing a particular symbolic representation strongly directs the behaviour of a theorem-prover. \cite[p.\ts96]{gonthier-ssreflect}
\end{quote}
So Ssreflect appears to sacrifice a degree of mathematical abstraction, though nobody can deny its success \cite{gonthier-4ct,gonthier-oot}. The Coquelicot analysis library similarly shies away from the full type system: 
\begin{quote}
	The Coq system comes with an axiomatization of standard real numbers and a library of theorems on real analysis. Unfortunately, \dots{} the definitions of integrals and derivatives are based on dependent types, which make them especially cumbersome to use in practice.'' \cite[p.\ts41]{boldo-coquelicot}
\end{quote}
In the sequel, we should be concerned with two questions:
\begin{itemize}
  \item whether simple type theory is sufficient for doing significant mathematics, and
  \item whether we can avoid getting locked into \textit{any} one formalism.
\end{itemize}
The latter, because it would be absurd to claim that any one formalism is all that we could ever need.

\subsection{Simple Type Theory}
Higher-order logic as implemented in proof assistants such as HOL Light \cite{hol-light-tutorial} and Isabelle/HOL \cite{isa-tutorial} borrows the syntax of types in the programming language~ML~\cite{paulson-ml2}. It provides
\begin{itemize}
\item \textit{atomic types},
in particular \isa{bool}, the type of truth values, and \isa{nat}, the type of natural numbers. 
\item \textit{function types}, denoted by \isa{$\tau_1$ \isasymRightarrow\ $\tau_2$}.
\item \textit{compound types}, such as \isa{$\tau$ list} for lists whose elements have type~$\tau$, similarly \isa{$\tau$ set} for typed sets. (Note the postfix notation.)
\item \textit{type variables},
  denoted by \isa{'a}, \isa{'b} etc. They give rise
  to polymorphic types like \isa{'a \isasymRightarrow~'a}, the type of the identity
  function.
\end{itemize}
Implicit in Church, and as noted above by de Bruijn, type variables and polymorphism must be included in the formalism implemented by a proof assistant. For already when we consider elementary operations such as the union of two sets, the type of the sets' elements is clearly a parameter and we obviously expect to have a single definition of union. Polymorphism makes that possible.

The terms of higher-order logic are precisely those of the typed $\lambda$-calculus: identifiers (which could be variables or constants), $\lambda$-abstractions and function applications. On this foundation a full predicate calculus is built, including equality. 
Note that while first-order logic regards terms and formulas as distinct syntactic categories, higher-order logic distinguishes between terms and formulas only in that the latter have type \isa{bool}. 

 \textit{Overloading} is the idea of using type information to disambiguate expressions. In a mathematical text, the expression $u\times v$ could stand for any number of things: $A\times B$ might be the Cartesian product of two sets, $G\times H$ the direct product of two groups and $m\times n$ the arithmetic product of two natural numbers. Most proof assistants make it possible to assign an operator such as $\times$ multiple meanings, according to the types of its operands. In view of the huge ambiguity found in mathematical notations--- consider for example $xy$, $f(x)$, $f(X)$, $df/dx$, $x^2y$, $\sin^2 y$---the possibility of overloading is a strong argument in favour of a typed formalism.

\subsection{Higher-Order Logic as a Basis for Mathematics}\label{sec:hol}

The formal deductive systems in HOL Light and Isabelle/HOL closely follow Church~\cite{church40}.  However, no significant applications can be tackled from this primitive starting point. It is first necessary to develop, at least, elementary theories of the natural numbers and lists (finite sequences). General principles of recursive/inductive definition of types, functions and sets are derived, by elaborate constructions, from the axioms. Even in the minimalistic HOL Light, this requires more than 10,000 lines of machine proofs; it requires much more in Isabelle, deriving exceptionally powerful recursion principles~\cite{blanchette-datatypes}.  This foundation is already sufficient for studying many problems in functional programming and hardware verification, even without negative integers.

To formalise analysis requires immensely more effort. It is necessary to develop the real numbers (as Cauchy sequences for example), but that is just the beginning. Basic topology including limits, continuity, derivatives, power series and the familiar transcendental functions must also be formalised. And all that is barely a foundation for university-level mathematics. In addition to the sheer bulk  of material that must be absorbed, there is the question of duplication. The process of formalisation gives rise to several number systems: natural numbers, integers, rationals, reals and complex numbers. This results in great duplication, with laws such as $x+0=x$ existing in five distinct forms. Overloading, by itself, doesn't solve this problem.

The need to reason about $n$-dimensional spaces threatens to introduce infinite duplication. Simple type theory does not allow dependent types, and yet the parameter~$n$ (the dimension) is surely a natural number. The theory of Euclidean spaces concerns $\mathbb{R}^n$ for any~$n$, and it might appear that such theorems cannot even be stated in higher-order logic. John Harrison found an ingenious solution~\cite{harrison-euclidean}: to represent the dimension by a type of the required cardinality. It is easy to define types in higher-order logic having any specified finite number of elements. Then $\mathbb{R}^n$ can be represented by the type $n\to \text{real}$, where the dimension $n$ is a type. Through polymorphism, $n$ can be a variable, and the existence of sum and product operations on types even allow basic arithmetic to be performed on dimensions. It must be admitted that things start to get ugly at this point. Other drawbacks include the need to write $\mathbb{R}$ as $\mathbb{R}^1$ in order to access topological results in the one-dimensional case. Nevertheless, this technique is flexible enough to support the rapidly expanding HOL Light multivariate analysis library, which at the moment covers much complex analysis and algebraic topology, including the Cauchy integral theorem, the prime number theorem, the Riemann mapping theorem, the Jordan curve theorem and much more. It is remarkable what can be accomplished with such a simple foundation.

It's important to recognise that John Harrison's approach is not the only one. An obvious alternative is to use polymorphism and explicit constraints (in the form of sets or predicates) to identify domains of interest. Harrison rejects this because
\begin{quotation}
  it seems disappointing that the type system then makes little useful contribution, for example in ‘automatically’ ensuring that one does not take dot products of vectors of different lengths or wire together words of different sizes. All the interesting work is done by set constraints, just as if we were using an untyped system like set theory. \cite[p.\ts115]{harrison-euclidean}
\end{quotation}
Isabelle/HOL provides a solution to this dilemma through an extension to higher-order logic: \textit{axiomatic type classes} \cite{wenzel-type}. This builds on the idea of polymorphism, which in its elementary form is merely a mechanism for type schemes: a definition or theorem involving type variables stands for all possible instances where types are substituted for the type variables. Polymorphism can be refined by introducing classes of types, allowing a type variable to be constrained by one or more type classes, and allowing a type to be substituted for a type variable only if it belongs to the appropriate classes. A type class is defined by specifying a suite of operations together with laws that they must satisfy, for example, a partial ordering with the operation $\le$ satisfying reflexivity, antisymmetry and transitivity or a ring with the operations 0, 1, $+$, $\times$ satisfying the usual axioms. The type class mechanism can express a wide variety of constraints using types themselves, addressing Harrison's objection quoted above. Type classes can also be extended and combined with great flexibility to create specification hierarchies: partial orderings, but also linear and well-founded orderings; rings, but also groups, integral domains, fields, as well as linearly ordered fields, et cetera. Type classes work equally well at specifying concepts from analysis such as topological spaces of various kinds, metric spaces and Euclidean spaces \cite{hoelzl-filters}.

Type classes also address the issue of duplication of laws such as $x+0=x$. That property is an axiom for the type class of groups, which is inherited by rings, fields, etc. As a new type is introduced (for example, the rationals), operations can be defined and proved to satisfy the axioms of some type class; that being done, the new type will be accepted as a member of that type class (for example, fields). This step can be repeated for other type classes (for example, linear orderings). At this point, it is possible to forget the explicit definitions (for example, addition of rational numbers) and refer to axioms of the type classes, such as $x+0=x$ . Type classes also allow operators such as~$+$ to be overloaded in a principled manner, because all of those definitions satisfy similar properties. Recall that overloading means assigning an operator multiple meanings, but when this is done through type classes, the multiple meanings will enjoy the same axiomatic properties, and a single type class axiom can replace many theorems~\cite{paulson-numerical}.

An intriguing aspect of type classes is the possibility of recursive definitions over the structure of types. For example, the lexicographic ordering $\le$ on type \isa{$\tau$~list} is defined in terms of $\le$ on type~$\tau$. But this introduces the question of circular definitions. More generally, it becomes clear that the introduction of type classes goes well beyond the na\"\i ve semantic foundation of simple type theory as a notation for a fragment of set theory. Recently, Kun{\v{c}}ar and Popescu \cite{kuncar-consistent-foundation} have published an analysis including sufficient conditions for overloaded constant definitions to be sound, along with a new semantics for higher-order logic with type classes. It remains much simpler than the semantics of any dependent type theory.

\subsection{A Personal Perspective}

In the spring of 1977, as a mathematics undergraduate at Caltech, I had the immense privilege of attending a lecture series on AUTOMATH given by de Bruijn himself, and of meeting him privately to discuss it. I studied much of the AUTOMATH literature, including Jutting's famous thesis \cite{jutting77} on the formalisation of Landau's \textit{Foundations of Analysis}. 


In the early 1980s, I encountered Martin-L\"of's type theory through the group at Chalmers University in Sweden. Again I was impressed with the possibilities of this theory, and devoted much of my early career to it.  I worked on the derivation of well-founded recursion in Martin-L\"of's type theory~\cite{paulson-cons}, and created Isabelle  originally as an implementation of this theory~\cite{paulson-natural}. Traces of this are still evident in everything from Isabelle's representation of syntax to the rules for $\Pi$, $\Sigma$ and $+$ constructions in Isabelle/ZF\@. The logic CTT (constructive type theory) is still distributed with Isabelle,\footnote{\url{http://isabelle.in.tum.de}}
including an automatic type checker and simplifier. 

My personal disenchantment with dependent type theories coincides with the decision to shift from extensional to intensional equality~\cite{nordstrom90}. This meant for example that $0+n=n$ and $n+0=n$ would henceforth be regarded as fundamentally different assertions, one an identity holding by definition and the other a mere equality proved by induction. Of course I was personally upset to see several years of work, along with Constable's Nuprl project~\cite{constable86}, suddenly put beyond the pale. But I also had the feeling that this decision had been imposed on the community rather than arising from a rational discussion. And I see the entire homotopy type theory effort as an attempt to make equality reasonable again.

\section{Example: Stereographic Projections}

An example will serve to demonstrate how mathematics can be formalised using the techniques described in \S\ref{sec:hol} above. We shall compare two formalisations of a theorem: the HOL Light original and the new version after translation to Isabelle/HOL using type classes.

\begin{figure}
\includegraphics[scale=.15]{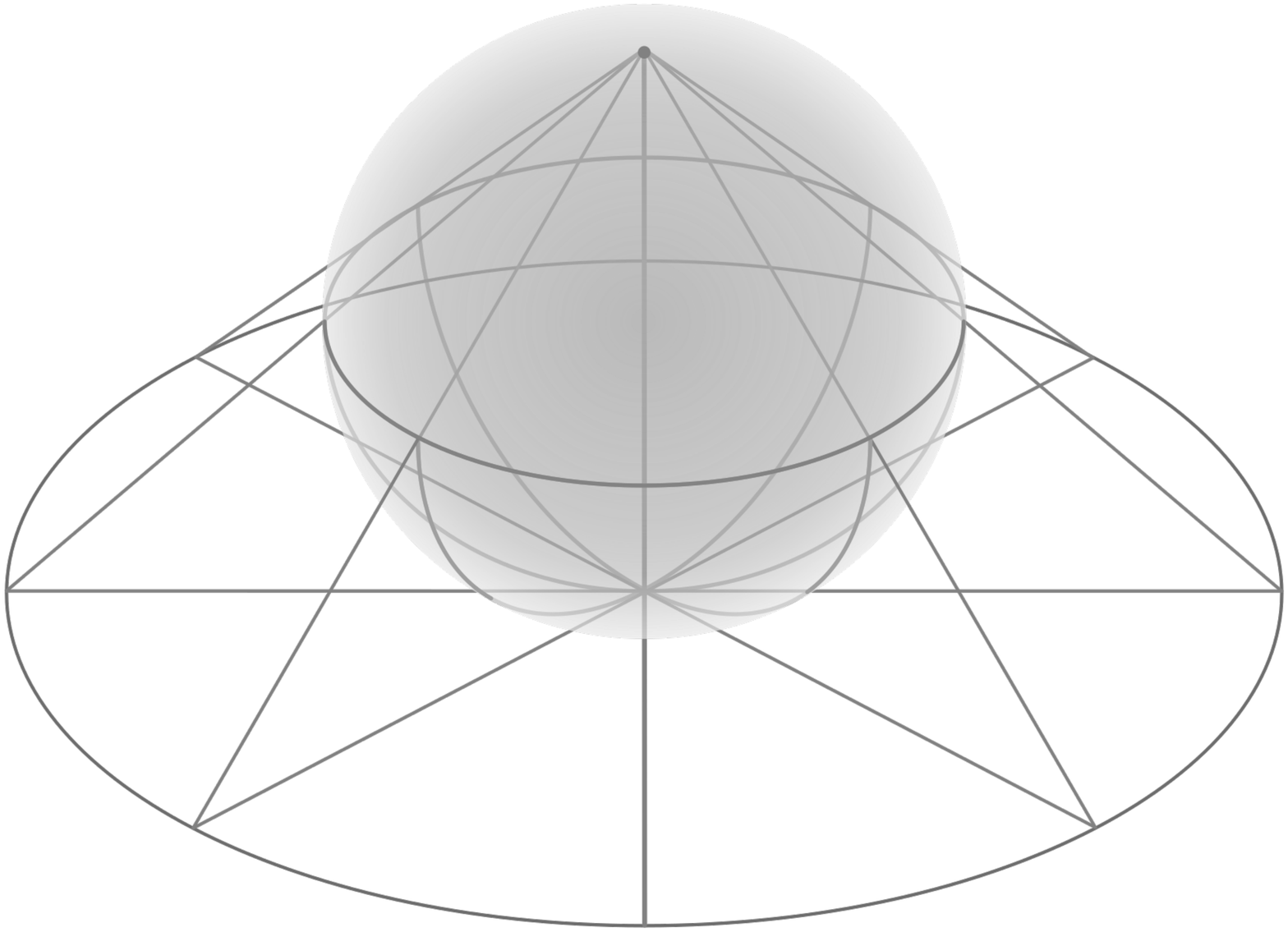}
\caption{3D illustration of a stereographic projection from the north pole onto a plane below the sphere}
\label{fig:stereo}  
\end{figure}

The theorem concerns stereographic projections, including the well-known special case of mapping a punctured\footnote{\textit{Punctured} means that one point is removed.} sphere onto a plane (Fig.\ts\ref{fig:stereo}). In fact, it holds under rather general conditions. In the two-dimensional case, a punctured circle is flattened onto a line. The line or plane is infinite, and points close to the puncture are mapped ``out towards infinity''. The theorem holds in higher dimensions with the sphere generalised to the surface of an $n$-dimensional convex bounded set and the plane generalised to an affine set of dimension $n-1$. The mappings are continuous bijections between the two sets: the sets are \textit{homeomorphic}.

 The theorem we shall examine is the generalisation of the case for the sphere to the case for a bounded convex set. The proof of this theorem is formalised in HOL Light\footnote{File \url{https://github.com/jrh13/hol-light/blob/master/Multivariate/paths.ml}}
 as shown in Fig.\ts\ref{fig:hol-sphere-thm}. At 51 lines, it is rather short for such proofs, which can be thousands of lines long.

\begin{figure}
\begin{scriptsize}\verbatimindent=0pt
\begin{verbatim}
let HOMEOMORPHIC_PUNCTURED_SPHERE_AFFINE_GEN = prove
 (`!s:real^N->bool t:real^M->bool a.
        convex s /\ bounded s /\ a IN relative_frontier s /\
        affine t /\ aff_dim s = aff_dim t + &1
        ==> (relative_frontier s DELETE a) homeomorphic t`,
  REPEAT GEN_TAC THEN ASM_CASES_TAC `s:real^N->bool = {}` THEN
  ASM_SIMP_TAC[AFF_DIM_EMPTY; AFF_DIM_GE; INT_ARITH
   `--(&1):int <= s ==> ~(--(&1) = s + &1)`] THEN
  MP_TAC(ISPECL [`(:real^N)`; `aff_dim(s:real^N->bool)`]
    CHOOSE_AFFINE_SUBSET) THEN REWRITE_TAC[SUBSET_UNIV] THEN
  REWRITE_TAC[AFF_DIM_GE; AFF_DIM_LE_UNIV; AFF_DIM_UNIV; AFFINE_UNIV] THEN
  DISCH_THEN(X_CHOOSE_THEN `t:real^N->bool` STRIP_ASSUME_TAC) THEN
  SUBGOAL_THEN `~(t:real^N->bool = {})` MP_TAC THENL
   [ASM_MESON_TAC[AFF_DIM_EQ_MINUS1]; ALL_TAC] THEN
  GEN_REWRITE_TAC LAND_CONV [GSYM MEMBER_NOT_EMPTY] THEN
  DISCH_THEN(X_CHOOSE_TAC `z:real^N`) THEN STRIP_TAC THEN
  MP_TAC(ISPECL
   [`s:real^N->bool`; `ball(z:real^N,&1) INTER t`]
        HOMEOMORPHIC_RELATIVE_FRONTIERS_CONVEX_BOUNDED_SETS) THEN
  MP_TAC(ISPECL [`t:real^N->bool`; `ball(z:real^N,&1)`]
        (ONCE_REWRITE_RULE[INTER_COMM] AFF_DIM_CONVEX_INTER_OPEN)) THEN
  MP_TAC(ISPECL [`ball(z:real^N,&1)`; `t:real^N->bool`]
        RELATIVE_FRONTIER_CONVEX_INTER_AFFINE) THEN
  ASM_SIMP_TAC[CONVEX_INTER; BOUNDED_INTER; BOUNDED_BALL; CONVEX_BALL;
               AFFINE_IMP_CONVEX; INTERIOR_OPEN; OPEN_BALL;
               FRONTIER_BALL; REAL_LT_01] THEN
  SUBGOAL_THEN `~(ball(z:real^N,&1) INTER t = {})` ASSUME_TAC THENL
   [REWRITE_TAC[GSYM MEMBER_NOT_EMPTY; IN_INTER] THEN
    EXISTS_TAC `z:real^N` THEN ASM_REWRITE_TAC[CENTRE_IN_BALL; REAL_LT_01];
    ASM_REWRITE_TAC[] THEN REPEAT(DISCH_THEN SUBST1_TAC) THEN SIMP_TAC[]] THEN
  REWRITE_TAC[homeomorphic; LEFT_IMP_EXISTS_THM] THEN
  MAP_EVERY X_GEN_TAC [`h:real^N->real^N`; `k:real^N->real^N`] THEN
  STRIP_TAC THEN REWRITE_TAC[GSYM homeomorphic] THEN
  TRANS_TAC HOMEOMORPHIC_TRANS
    `(sphere(z,&1) INTER t) DELETE (h:real^N->real^N) a` THEN
  CONJ_TAC THENL
   [REWRITE_TAC[homeomorphic] THEN
    MAP_EVERY EXISTS_TAC [`h:real^N->real^N`; `k:real^N->real^N`] THEN
    FIRST_X_ASSUM(MP_TAC o GEN_REWRITE_RULE I [HOMEOMORPHISM]) THEN
    REWRITE_TAC[HOMEOMORPHISM] THEN STRIP_TAC THEN REPEAT CONJ_TAC THENL
     [ASM_MESON_TAC[CONTINUOUS_ON_SUBSET; DELETE_SUBSET];
      ASM SET_TAC[];
      ASM_MESON_TAC[CONTINUOUS_ON_SUBSET; DELETE_SUBSET];
      ASM SET_TAC[];
      ASM SET_TAC[];
      ASM SET_TAC[]];
    MATCH_MP_TAC HOMEOMORPHIC_PUNCTURED_AFFINE_SPHERE_AFFINE THEN
    ASM_REWRITE_TAC[REAL_LT_01; GSYM IN_INTER] THEN
    FIRST_X_ASSUM(MP_TAC o GEN_REWRITE_RULE I [HOMEOMORPHISM]) THEN
    ASM SET_TAC[]]);;	
\end{verbatim}
\end{scriptsize}
\label{fig:hol-sphere-thm}  
\end{figure}

The HOL Light proof begins with the statement of the desired theorem. We see logical syntax coded as ASCII characters: \texttt{!} = $\forall$ and \verb|/\| = $\land$.  Moreover, the \texttt{DELETE} operator refers to the removal of a set element ($S-\{a\}$). Words such as \texttt{convex} and \texttt{bounded} denote predicates defined elsewhere. Infix syntax is available, as in the symbol \texttt{homeomorphic}. We see John Harrison's representation of $\mathbb{R}^n$ in the type \verb|real^N->bool| and in particular, \verb|!s:real^N->bool| abbreviates ``for all $s\subseteq \mathbb{R}^N$''. Note that the constraint on the dimensions is expressed through the concept of affine dimension rather than some constraint on \texttt{M} and~\texttt{N}\@. This statement is legible enough, and yet the notation leaves much to be desired, for example in the necessity to write \verb|&1| (the ampersand converting the natural number~1 into an integer).

\begin{small}
\begin{verbatim}
!s:real^N->bool t:real^M->bool a.
     convex s /\ bounded s /\ a IN relative_frontier s /\
     affine t /\ aff_dim s = aff_dim t + &1
     ==> (relative_frontier s DELETE a) homeomorphic t	
\end{verbatim}
\end{small}

We have to admit that the proof itself is unintelligible. Even a HOL Light user can only spot small clues in the proof text, such as the case analysis on whether the set $s$ is empty or not, which we see in the first line, or the references to previous lemmas. If we look carefully, we might notice intermediate statements being proved, such as

\begin{small}
\begin{verbatim}
~(t:real^N->bool = {})
\end{verbatim}
\end{small}

or

\begin{small}
\begin{verbatim}
~(ball(z:real^N,&1) INTER t = {})
\end{verbatim}
\end{small}

\noindent
though in the latter case it is unclear what $z$ is. The formal proof consists of program code, written in a general-purpose programming language (OCaml) equipped with a library of proof procedures and supporting functions, for that is what HOL Light is. A HOL Light proof is constructed by calling the its proof primitives at the OCaml command line, but one could type in any desired OCaml code.  Users sometimes write such code in order to extend the functionality of HOL Light. Even if their code is incorrect,%
\footnote{Malicious code is another matter. In HOL Light, one can use ocaml's \texttt{String.set} primitive to replace T (true) by F\@. Given the variety of loopholes in programming languages and systems, not to mention notational trickery, we must be content with defences against mere incompetence.}
 they cannot cause HOL Light to generate false theorems. All LCF-style proof assistants employ a similar kernel architecture.
\begin{figure}
\includegraphics[scale=.7]{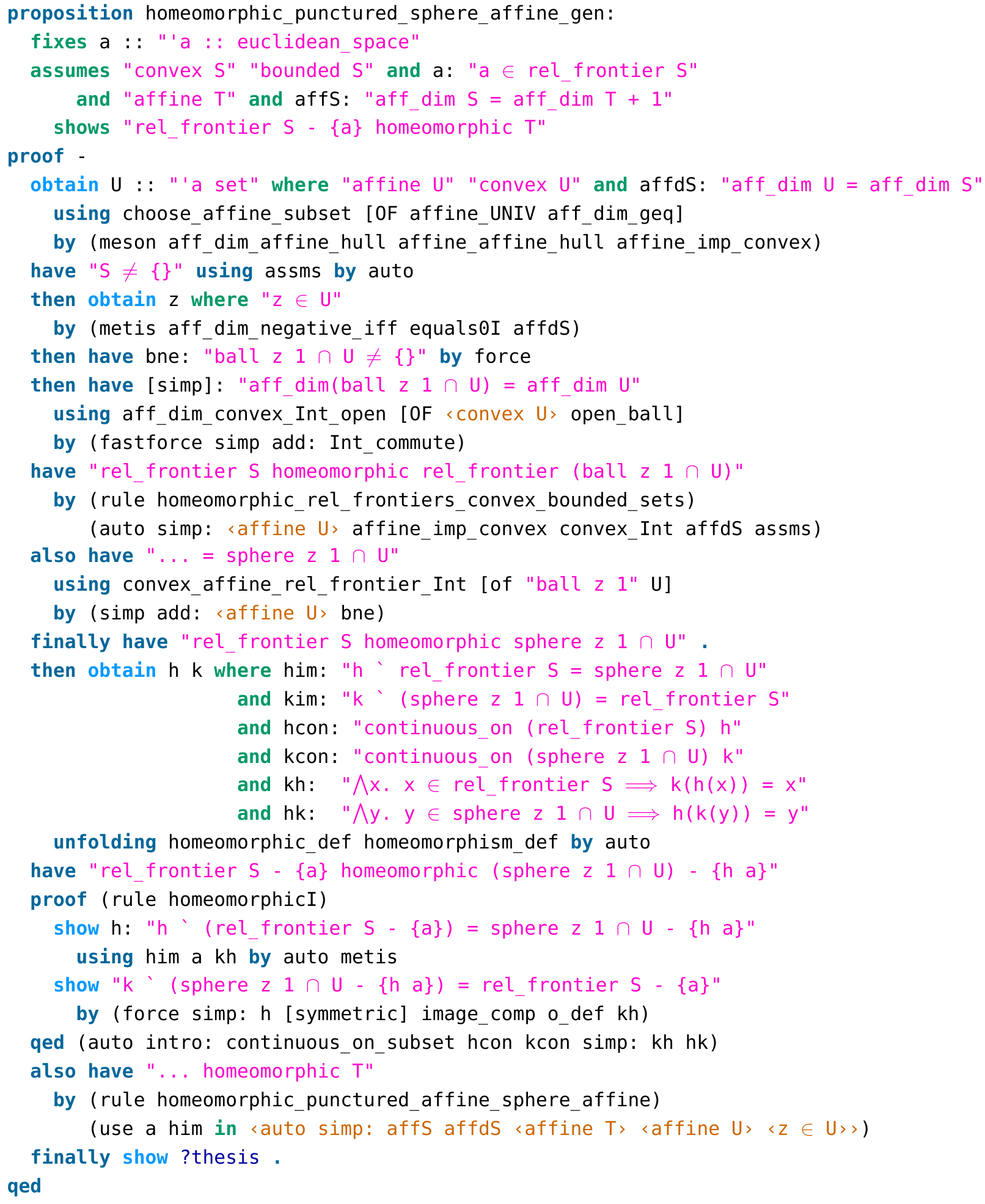}
\caption{The stereographic projection theorem in Isabelle/HOL}
\label{fig:isa-sphere-thm}  
\end{figure}

In recent years, I have been embarked on a project to translate the most fundamental results of the HOL Light multivariate analysis library into Isabelle. The original motivation was to obtain the Cauchy integral theorem~\cite{harrison-complex}, which is the gateway to the prime number theorem~\cite{harrison-pnt} among many other results. I was in a unique position to carry out this work as a developer of both Isabelle and HOL\@. The HOL family of provers descends from my early work on LCF \cite{paulson87}, and in particular the proof tactic language, which is perfectly preserved in HOL Light. The 51 lines of HOL Light presented above are among the several tens of thousands that I have translated Isabelle/HOL\@. Figure~\ref{fig:isa-sphere-thm} presents my version of the HOL Light proof above, as shown in a running Isabelle session. Proof documents can also be typeset with the help of \LaTeX, but here we have colour to distinguish the various syntactic elements of the proof: keywords, local variables, global variables, constants, etc.

The theorem statement resembles the HOL Light one but uses the Isabelle \textbf{fixes} / \textbf{assumes} / \textbf{shows} keywords to declare the premises and conclusion. (It is typical Isabelle usage to minimise the use of explicit logical connectives in theorem statements.) Harrison's construction \verb|real^N| isn't used here; instead the variable $a$ is declared to belong to some arbitrary Euclidean space. An advantage of this approach is that types such as \isa{real} and \isa{complex} can be proved to be Euclidean spaces despite not having the explicit form \verb|real^N|.

The proof is written in the Isar structured language, and much of it is legible. An affine set~$U$ is somehow obtained, with the same dimension as $S$, which we note to be nonempty, therefore obtaining some element $z\in U$. Then we obtain a homeomorphism between \isa{rel\_frontier S} and \isa{sphere z 1 \isasyminter\ U}, using a previous result.\footnote{Because the HOL Light libraries were ported en masse, corresponding theorems generally have similar names and forms.}
 Then an element is removed from both sides, yielding a new homeomorphism, which is chained with the homeomorphism theorem for the sphere to yield the final result. And thus we get an idea how the special case for a punctured sphere intersected with an affine set can be generalised to the present result.
 
 The Isar proof language~\cite{wenzel-isabelle/isar}, inspired by that of the Mizar system~\cite{trybulec-features}, encourages the explicit statement of intermediate results and obtained quantities. The notation also benefits from Isabelle's use of mathematical symbols, and a further benefit of type classes is that a number like 1 belongs to all numeric types without explicit conversion between them.
 
\section{Discussion and Conclusions}
 
The HOL Light and Isabelle proofs illustrate how mathematical reasoning is done in simple type theory. They also show what mathematics looks like in these systems. The Isabelle proof demonstrates that simple type theory can deliver a degree of legibility, though the syntax is a far cry from normal mathematics. The greater expressiveness of dependent type theories has not given them any advantage in the domain of analysis: the leading development~\cite{boldo-coquelicot} is not constructive and downgrades the role of dependent types.

As I have remarked elsewhere~\cite{paulson-computational-logic}, every formal calculus is ultimately a prison. It will do some things well, other things badly and many other things not at all. Mathematics write their proofs using a combination of prose and beautiful but highly ambiguous notations, such as 
${\partial^2z}/{\partial x^2} = y^2/x^2 + e^xy\cos y$. Formal proofs are code and look like it, even if they are allowed to contain special symbols and Greek letters. The various interpretations of anomalous expressions such as $x/0$ are also foundational, and each formalism must adopt a clear position when one might prefer a little ambiguity. (Both HOL Light and Isabelle define $x/0=0$, which some people find shocking.)  Develop our proof tools as we may, such issues will never go away. But if past generations of mathematicians could get used to REDUCE and FORTRAN, they can get used to this.

The importance of legibility can hardly be overstated. A legible proof is more likely to convince a sceptical mathematician: somebody who doesn't trust a complex software system, especially if it says $x/0=0$. While much research has gone into the verification of proof procedures \cite{kumar-self-formalisation,schlichtkrull-resolution}, all such work requires trusting similar software. But a mathematician may believe a specific formal proof if it can be inspected directly, breaking this vicious cycle. Ideally, the mathematician would then gain the confidence to construct new formal proofs, possibly reusing parts of other proofs. Legibility is crucial for this.

These examples, and the great proof translation effort from which they were taken, have much to say about the process of porting mathematics from one system to another. Users of one system frequently envy the libraries of a rival system. There has been much progress on translating proofs automatically \cite{kaliszyk-scalable-lcf,obua-importing-hol}, but such techniques are seldom used. Automatic translation typically works via a proof kernel that has been modified to generate a trace, so it starts with an extremely low-level proof. Such an approach can never deliver legible proofs, only a set of mechanically verified assertions. Manual translation, while immensely more laborious, yields real proofs and allows the statements of theorems to be generalised to take advantage of Isabelle/HOL's type classes.

All existing proof translation techniques work by emulating one calculus within another at the level of primitive inferences. Could proofs instead be translated at the level of a mathematical argument? I was able to port many proofs that I did not understand: despite the huge differences between the two proof languages, it was usually possible to guess what had to be proved from the HOL Light text, along with many key reasoning steps. Isabelle's automation was generally able to fill the gaps. This suggests that in the future,  if we start with structured proofs, they could be translated to similarly structured proofs for a new system. If the new system supports strong automation (and it must!), the translation process could be driven by the text alone, even if the old system was no longer available. The main difficulty would be to translate statements from the old system so that they look natural in the new one. 

The huge labour involved in creating a library of formalised mathematics is not in vain if the library can easily be moved on. The question ``is simple type theory the right foundation for mathematics?'' then becomes irrelevant. Let's give G\"odel the last word (italics his):
\begin{quote}
	Thus we are led to conclude that, although everything mathematical is formalisable, it is nevertheless impossible to formalise all of mathematics in a \textit{single} formal system, a fact that intuitionism has asserted all along. \cite[p.\ts389]{goedel35b}
\end{quote}

\begin{acknowledgement}
Dedicated to Michael J C Gordon FRS, 1948--2017.
The development of HOL and Isabelle has been supported by numerous EPSRC grants. The ERC project ALEXANDRIA supports continued work on the topic of this paper. 
Many thanks to Jeremy Avigad, Johannes H\"olzl, Andrew Pitts and the anonymous referee for their comments.
\end{acknowledgement}

\bibliographystyle{abbrv}
\bibliography{string,atp,general,isabelle,theory,funprog,crossref}
\end{document}